\input{aipcheck}

\documentclass[
    ,final            
  ]
  {aipproc}

\layoutstyle{6x9}

\IfFileExists{srcltx.sty}{\usepackage[active]{srcltx}}

\begin{document}

\title{Past Galactic GRBs, and the origin and composition of ultrahigh-energy cosmic rays
}


\author{Alexander Kusenko}{
  address={Department of Physics and Astronomy, University of California, Los
Angeles, CA 90095, USA\\ and \\
IPMU, University of Tokyo, Kashiwa, Chiba 277-8568, Japan}
}

\begin{abstract}
Recent results from the Pierre Auger Observatory show energy dependent chemical 
composition of ultrahigh-energy cosmic rays (UHECR) with a growing fraction of heavy 
elements at high energies.  This points to a non-negligible contribution of the Galactic sources, 
such as past GRBs and other rare but powerful stellar explosions in the Milky Way. The effects of diffusion in the 
Galactic magnetic fields alter the observed composition and render the flux of UHECR isotropic, 
up to a few per cent anisotropy in the direction of the Galactic Center, as well as some small-scale anisotropy 
with ``hot spots'' due to the locations of the most recent/closest bursts. 
\end{abstract}

\maketitle


Composition of ultrahigh-energy cosmic rays (UHECR) measured by the Pierre Auger Observatory (PAO) shows a steady increase of the mean nuclear mass with energy between 2~EeV and 35~EeV~\cite{Abraham:2009dsa,Abraham:2010yv}. This unexpected result is difficult to reconcile with the usually assumed extragalactic origin of UHECR.  The lack of plausible sources in the Milky Way and the lack of Galactocentric anisotropy of the arrival directions of UHECR are usually presented as evidence for extragalactic origin of UHECR above 10$^{18}$~eV. However, it was recently shown that, if the cosmic rays of 10$^{18}-10^{19}$~eV are nuclei produced in the Milky Way, the effects of diffusion in turbulent Galactic micro-Gauss magnetic fields can explain both the change in composition and the approximate anisotropy~\cite{Calvez:2010uh}.  
As for the plausible sources, there is a growing evidence that long GRBs are caused by a relatively rare type of supernovae, while the short GRBs probably result from the coalescence of neutron stars and/or black holes. Compact star mergers undoubtedly take place in the Milky Way, and therefore short GRBs should occur in our Galaxy. Although there is some correlation of long GRBs with star-forming metal-poor galaxies~\cite{Fruchter:2006py}, many long GRBs are observed in high-metallicity galaxies as well~\cite{Savaglio:2006xe,CastroTirado:2007tn,Levesque:2010rn}, and therefore one expects that long GRBs should occur in the Milky Way.  Less powerful hypernovae, too weak to produce a GRB, but can still accelerate UHECR~\cite{Wang:2007ya}, with a substantial fraction of nuclei~\cite{Wang:2007xj,Murase:2008mr}.  If the observed cosmic rays originate past explosions in our own Galaxy, PAO results have a straightforward explanation~\cite{Calvez:2010uh}.  

GRBs have been proposed as the sources of extragalactic UHECR~\cite{Waxman:1995vg,Vietri:1995hs,Murase:2008mr}, and they have also been considered as possible Galactic sources~\cite{Dermer:2005uk,Biermann:2003bt,Biermann:2004hi}. It is believed that GRBs, hypernovae, or other stellar events capable of producing UHECR could have happened in the Milky Way at the rate of one per  $t_{GRB}\sim 10^{4}-10^{5}$ years \cite{Schmidt:1999iw,Frail:2001qp,Furlanetto:2002sb,Perna:2003bi}. Such events have been linked to the observations of INTEGRAL, Fermi and PAMELA~\cite{Bertone:2004ek,Parizot:2004ph,Ioka:2008cv,Calvez:2010fd}. As illustrated in Fig.~1, diffusion depends on rigidity, and, therefore, the observed composition can be altered by diffusion~\cite{Wick:2003ex,Calvez:2010uh}.

\begin{figure}
   \includegraphics[width=.7\textwidth]{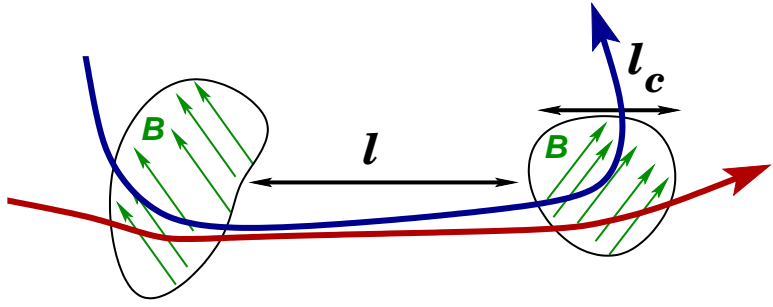}
  \caption{For each species, there is a critical energy $E_{0,i}$ for which the Larmor radius $R_i$ is equal to the magnetic coherence length $l_c$. For $E\ll E_{0,i}$, the mean free path of the diffusing particle is $l \sim l_0$, and $D_i(E)= l_c/3$.  For $E\gg E_{0,i}$, the particle is deflected only by a small angle $\theta\sim l_0/R_i$, and, after $k$ deflections, the mean deflection angle squared is $\bar{\theta^2}\sim k (l_0/R_i)^2$.  The corresponding diffusion coefficient is $D_i(E) \propto ( \frac{E}{E_{0,i}})^2 , \ {\rm for} \ E \gg  E_{0,i}$.}
\end{figure}

Changes in composition due to a magnetic fields have  been discussed in connection with the spectral ``knee''~\cite{Wick:2003ex}, and also for a transient source~\cite{Kotera:2009ms}. A simple model of diffusion~\cite{Calvez:2010uh} shows how the diffusion coefficient affects the observed spectrum of the species ``$i$''.  If all species are produced with the same spectrum $n_i^{(\rm src)}=n_0^{(\rm src)}\propto E^{-\gamma} $ at the source, their observed spectra are altered by the energy dependent diffusion and by the  trapping in the Galactic fields, so that, instead of $n_i^{(\rm src)}=n_0$, one obtains 
\begin{equation}
n_i(E,r) = \frac{Q_0}{4\pi r\, D_i(E)} \left ( \frac{E_0}{E} \right )^\gamma.
\label{solution_GC}
\end{equation}
Hence, the composition becomes energy dependent.  Indeed, at critical energy $E_{0,i}$, which is different for each nucleus, the solution (\ref{solution_GC}) changes from $\propto E^{-\gamma} $ to $\propto E^{-\gamma-2} $ because of the change in $D_i(E)$, as discussed in the caption of Fig.~1.  Since the change occurs at a rigidity-dependent critical energy $E_{0,i}=e E_0 Z_i$, the larger nuclei lag behind the lighter nuclei in terms of the critical energy and the change in slope.  If protons dominate for $E<E_0$, their flux drops dramatically for $E>E_0$, and the heavier nuclei dominate the flux.  The higher $Z_i$, the higher is the energy at which the species experiences a drop in flux.

One can also understand the change in composition by considering the time of diffusion across the halo is $t_i \sim R^2/D_i$.  The longer the particle remains in the halo, the higher is the probability of its detection.   At higher energies, the magnetic field's ability to delay the passage of the particle diminishes, and the density of such particles drops precipitously for $E>E_{0,i}$.  Since $E_i$ is proportional to the  electric charge, the drop in the flux occurs at different energies for different species.

\begin{figure}
  \includegraphics[height=.3\textheight]{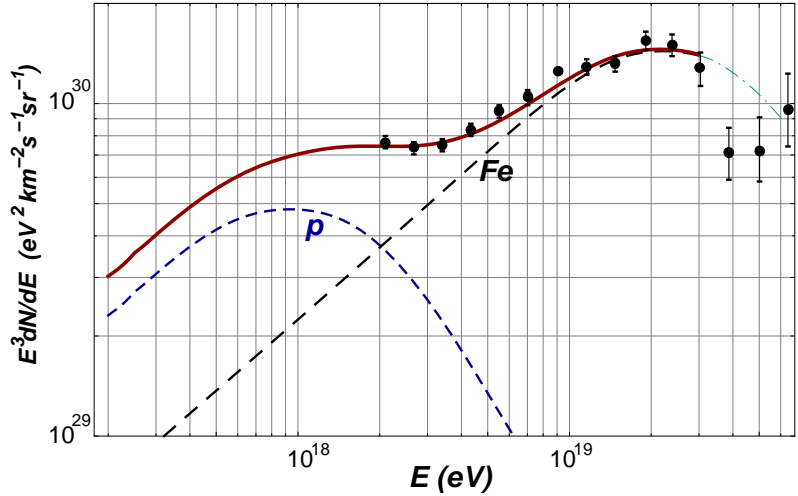}
  \caption{UHECR spectra according to Calvez {\em et al.} (shown here for values of parameters that differ from those in Ref.~\cite{Calvez:2010uh}) from 
Galactic sources assumed to produce 75\% protons and 25\% iron, with identical spectra $\propto  E^{-2.3}$. The source distribution traces the distribution of stars in the Galaxy.  The magnetic field was assumed to be $\sim 10\mu$G, coherent over $l_0=100$~pc domains. The overall power and the iron fraction were adjusted to fit Pierre Auger data points~\cite{Abraham:2009wk} (shown). 
}
\end{figure}

The model of Ref.~\cite{Calvez:2010uh} gives a qualitative description of the data. To reproduce the data more accurately, it must be improved.  First, one should use a more realistic source population model.  Second, one should include the coherent component of the Galactic magnetic field.  Third, one should not assume that UHECR comprise only two types of particles, and one should include a realistic distribution of nuclei. Finally, one should include the extragalactic component of UHECR produced by distant sources, such as active galactic nuclei (AGN) and GRBs (outside the Milky Way).  A recent realization that very high energy gamma rays observed by Cherenkov telescopes from distant blazars are likely to be secondary photons produced in cosmic ray interactions along the line of sight lends further support to the assumption that cosmic rays are copiously produced in AGN jets~\cite{Essey:2009zg,Essey:2009ju}. For energies $E>3\times 10^{19}$~eV, the energy losses due to photodisintegration, pion production, pair production and interactions with interstellar medium become important and must be included. The propagation distance in the Galaxy exceeds 10~Mpc, so that the Galactic component should exhibit an analog of GZK suppression in the spectrum.  The extragalactic propagation can also affect the composition around $10^{18}$~eV~\cite{Hill:1983mk}.

Galactocentric anisotropy for a source distribution that traces the stellar counts in the Milky Way is small~\cite{Calvez:2010uh}. Although the anisotropy in protons is large at high energies, their contribution to the total flux is small, so the total anisotropy was found to be $<10\%$, consistent with the observations.  The latest GRBs do not introduce a large degree of anisotropy, as it would be in the case of UHE protons, but they can create ``hot spots'' and clusters of events.

Our model~\cite{Calvez:2010uh} makes an interesting prediction for the highest-energy cosmic rays.  Just as the protons of the highest energies escape from our Galaxy, they should escape from the host galaxies of remote sources, such as AGNs. Therefore, UHECR with $E>3\times 10^{19}$~eV should correlate with the extragalactic sources.  Moreover, these UHECR should be protons, not heavy nuclei, since the nuclei are trapped in the host galaxies.  If and when the data will allow one to determine composition on a case-by-case basis, one can separate $E>3\times 10^{19}$~eV events into protons and nuclei and observe that the protons correlate with the nearby AGN.  This prediction is one of the non-trivial tests of our model: at the highest energies the proton fraction should exist and should correlate with known astrophysical sources.

\begin{theacknowledgments}

 The author thanks A.~Calvez and S.~Nagataki for fruitful collaboration that produced the results reviewed above. This work was supported  by DOE Grant DE-FG03-91ER40662 and NASA ATP Grant NNX08AL48G.
\end{theacknowledgments}



\bibliographystyle{aipproc}   

\bibliography{grb}

\IfFileExists{\jobname.bbl}{}
 {\typeout{}
  \typeout{******************************************}
  \typeout{** Please run "bibtex \jobname" to optain}
  \typeout{** the bibliography and then re-run LaTeX}
  \typeout{** twice to fix the references!}
  \typeout{******************************************}
  \typeout{}
 }

\end{document}